\documentstyle[aps,prd,epsfig]{revtex}

\tightenlines

\begin{document}
\title{Analyticity Relations at High Energies and Forward
Scattering\thanks{LISHEP
2002 - Session C: Workshop On Diffractive Physics - February 4 - 8, 2002, Rio
de Janeiro, RJ, Brazil}}
\author{R.F. \'Avila$^{a}$, E.G.S. Luna$^{b}$, M.J. Menon$^{b}$} 
\address{$^a$Instituto de Matem\'atica,
Estat\'{\i}stica e Computa\c c\~ao Cient\'{\i}fica \\
\small\it $^b$Instituto de F\'{\i}sica Gleb Wataghin \\
Universidade Estadual de Campinas - UNICAMP\\
13083-970 Campinas, SP, Brazil} 

\maketitle

\begin{abstract}

Making use of a derivative dispersion approach, we investigate
the behavior of the the total cross section and the $\rho$ parameter for $pp$
and $\overline{p}p$ scattering from accelerator to cosmic ray energies. The
discrepancies in the cosmic ray information is treated through the definition of
two different ensembles of data. Simultaneous and individual fits to the above
quantities through the Donnachie-Landshof and Kang-Nicolescu parametrizations 
allows to infer an upper bound for the intercept of the soft Pomeron and
also show that the data investigated do not favor the Odderon hypothesis.
\end{abstract}

\section{Introduction}

Forward quantities (total cross section, $\rho$, slope and
curvature parameters) are fundamental quantities in the investigation of 
hadron-hadron scattering. In particular, total cross section and $\rho$ are
expressed in terms of the real and imaginary parts of the forward scattering
amplitude by \cite{matthiae,bc}

\begin{eqnarray} 
\sigma_{tot}(s) = \frac{Im F(s,t=0)}{s}, \qquad 
\rho(s) = \frac{Re F(s,t=0)}{Im F(s,t=0)},
\end{eqnarray} 
where, $\sqrt s$ is the center of mass energy and $t$ the four-momentum
transfer squared.

 The theoretical description of these quantities
as function of the energy constitute, presently, a great challenge.
Among the variety of approaches, the analytical
models are characterized by suitable analytic parametrizations to both
quantities and the use of dispersion relations connecting real and imaginary
parts of the amplitude (Principle of Analyticity) . Also, particle-particle and
particle-antiparticle scattering are simultaneously investigated by combining
even and odd amplitudes (Principle of Crossing Symmetry).

Presently, the highest energies reached in accelerator experiments correspond
to antiproton-proton scattering, $\sqrt s =1.8$ TeV and this value
is nearly thirty times the highest energy reached in the case of
proton-proton scattering, $\sqrt s = 62.5$ GeV. These differences are the
responsible for the difficulties present in the simultaneous
investigation of both reactions, meanly in the case of the behaviors as
function of the energy and asymptotic limits.

In this communication we treat $pp$ and $\overline{p}p$ forward
scattering through analytical models and the central point is a critical
investigation based on two observations:

\begin{description}
\item[1.] The total cross section is an experimental quantity,
basically determined by counting the number of elastic and inelastic events in
the scattering process. On the other hand, $\rho$ is a free fit
parameter, obtained through fits to the differential cross section data in the
region of interference Coulomb - nuclear \cite{matthiae,bc}. Therefore, the
values of $\rho$ extracted from experiments are model dependent. We conclude
that $\sigma_{tot}$ and $\rho$ have different status as physical quantities and
this puts limitations in the simultaneous investigation of both quantities in
constrained approaches, such as dispersion relations.
\item[2.] Beyond the energy region of the accelerators there are
experimental information on $pp$ total cross section from cosmic ray
experiments. Despite of the high error bars in the numerical results
and also some discrepancies, they are usually shown as a support to several
model and analysis predictions.
\end{description}

The first observation suggests the investigation of the effect of
simultaneous and independent fits to $\sigma_{tot}(s)$ and $\rho(s)$
and the second, the determination of  possible statistical bounds  on the raise
of these quantities, due to all the cosmic ray information presently available
(taking account of the large error bars). These are the points we are
interested in.
The basic procedure is the choice of suitable parametrizations for
$\sigma_{tot}$ with different asymptotic behaviors and the use of analyticity
(dispersion) relations in order to connect $\sigma_{tot}(s)$ and $\rho(s)$.
Observation 1 is treated through individual and simultaneous fits to both
quantities and observation 2 by selecting different ensembles of data
based on distinct cosmic ray information, as will be explained in detail.
The paper is organized as follows. In Sec. II we present the experimental data
investigated and discuss the discrepancies concerning cosmic ray information.
In Sec. III we review the main formulas in the dispersion relation approach
and introduce two different parametrizations for the total cross section,
namely the Donnachie-Landshof and the Kang-Nicolescu models. The results of
all the fits are presented in Sec. IV and the conclusions and some final
remarks are the contents of Sec. V.

\section{Experimental data and Ensembles}

\subsection{\bf Accelerator Data}

In the case of accelerator experiments, data on $\sigma_{tot}$ 
from $pp$ and $\overline{p}p$  total cross sections and also estimations of the
$\rho$ parameter have been accumulated for a long time.
The data sets, compiled and analyzed by the Particle Data
Group, have become a standard reference and are available (also readable
files) at {\it http://pdg.lbl.gov}. Recent analysis has shown
that  general fits
are stable for $\sqrt s$ above $\sim$ 9 GeV \cite{cudell}. For this reason in
this work we make use of these sets for energies above $10$ GeV and in our
analysis the statistic and systematic errors were added linearly. The total
cross section data are shown in Fig. 1.

\subsection{\bf Cosmic ray information}

The extraction of the proton-proton total cross section from cosmic ray
experiments is based on the determination of the proton-air production
cross section from analysis of extensive air showers. Detailed review on the
subtle involved may be found in References \cite{gaisser} and \cite{engel}.
Here we only recall the two main steps, stressing the model dependence
involved.

The first step concerns the determination of the proton-air production cross
section, namely, the ``inelastic cross section in which at least one new
hadron is produced in addition to nuclear fragments'' \cite{engel}. This is
obtained by the formulas

\begin{eqnarray}  
\sigma_{p-air}^{prod}(mb) = \frac{2.4 \times 10^4}{\lambda_{p-air}(g/cm^2)},
\qquad \lambda_{p-air}(g/cm^2) = \frac{\lambda_{att}}{k}, 
\end{eqnarray} 
where, $\lambda_{p-air}$ is the interaction length of protons in the
atmosphere, $\lambda_{att}$ the shower attenuation length and  
$k$ is a measure of the dissipation of the energy through the shower. The
$\lambda_{att}$ is a experimental quantity determined through the
$\chi_{max}$ Attenuation Method or the Zenith Angle Attenuation Technique
\cite{sokolsky}. On the other hand, the parameter $k$ is obtained through
Monte Carlo simulation and so is strongly model dependent \cite{engel}.

In a second step $\sigma_{pp}^{tot}$ is obtained from $\sigma_{p-air}^{prod}$
through the Multiple Diffraction Formalism by Glauber and Matthiae \cite{gm}
and taking account of the screening correction ($\Delta \sigma$):

\begin{eqnarray}  
\sigma_{p-air}^{tot} = \sigma_{p-air}^{el}  + 
\sigma_{p-air}^{prod} + \sigma_{p-air}^{q-el} + \sigma_{p-air}^{diff} +
\Delta \sigma \nonumber
\end{eqnarray}

Two important inputs at this point are the nucleon distribution function and the
slope parameter $B$, necessary in the parametrization of the proton-proton
scattering amplitude:

\begin{eqnarray}  
F_{pp}(s,q) \propto \exp \{ \frac{-B(s) q^2}{2}\},
\end{eqnarray} 
where $q^2$ is the four-momentum transfer squared.

The results presently available from cosmic ray experiments cover the region
$\sqrt s \sim 6 - 40$ TeV and are
characterized by discrepancies, mainly due to the model dependence of $k$ and
$B(s)$. However, we can classify these results according to the degree of
consistence of the analysis and, simultaneously, by value of the total cross
section extracted, as explained in what follows. 

From one hand the results usually quoted in the literature concern those
obtained by the Fly's Eye Collaboration in 1984 \cite{fly} and Akeno (Agasa)
Collaboration in 1993 \cite{akeno}.
In the first case, the authors used $k = 1.6$ and a relation of proportionality
between $B$ and $\sigma_{pp}^{tot}$ as predicted by the Geometrical Scaling
model \cite{jorge}. At $\sqrt s = 30$ TeV the cross sections results are
\cite{fly}:

\begin{eqnarray}
\sigma_{p-air}^{prod} = 540 \pm 50 {\rm \ mb } \qquad 
\sigma_{pp}^{tot} = 122 \pm 11 {\rm \ mb} \nonumber 
\end{eqnarray} 
The Akeno Collaboration used $k = 1.5$ and the Durand and Pi method
\cite{dp1,dp2} to extract the proton-proton cross sections in the region $\sqrt
s \sim 6 - 20$ TeV. The results are displayed in Fig. 1 together with the  Fly's
Eye value at $\sqrt s = 30$ TeV.

On the other hand, in 1987 Gaisser, Sukhatme, Yodh (GSY) making use of the
Fly's Eye result for $\sigma_{p-air}^{prod}$ and the Chou - Yang prescription
for $B(s)$ obtained at $\sqrt s = 40$ TeV \cite{gsy}:
\begin{eqnarray}
\sigma_{pp}^{tot} = 175_{-27}^{+40}  
\end{eqnarray}
In 1993, Nikolaev claimed that the Akeno results should be corrected
in order to take account of the differences between absorption and
inelastic cross sections, leading to an increase of the published results by
$\approx 30$ mb \cite{niko}.

All these cosmic ray estimates together with the accelerator results are
displayed in Fig 1 and show that, despite of the large error bars, we can
identify two distinct set of estimations: one represented by the results of
the Fly's Eye Collaboration together with those by the Akeno
Collaboration; the other by the results of Gaisser, Sukhatme, Yodh
with those by Nikolaev. Taken separately these two sets suggest different
behaviors for the increase of the total cross section.

At this point it should be stressed that the Fly's Eye Collaboration
used the Geometrical Scaling hypothesis, which is violated even at the
collider energy and so this result is certainly wrong. Durand and Pi
asserted in Ref. \cite{dp2} that their results presented in \cite{dp1},
and used by the Akeno Collaboration,
should be disregarded due to a wrong approximation. On the other
hand, to our knowledge, the results by Gaisser, Sukhatme, Yodh
and Nikolaev have never been critized in the literature. But we 
must yet take account of the model dependence involved as discussed in
\cite{gaisser,engel}.  

For the above reasons we shall 	quantitatively investigate the
statistical determination of possible lower or upper limits in the increase of
the total cross section, associated with the results by Akeno - Fly's Eye
and Nikolaev - GSY, respectively. To this end
we consider two ensembles of data, above $10$ GeV, defined by: 

\begin{description}
\item[Ensemble I:] $\overline{p}p$ accelerator data and $pp$ accelerator data +
Akeno + Fly's Eye.  
\item[Ensemble II:] $\overline{p}p$ accelerator data and 
 $pp$  accelerator data + Nikolaev + GSY. 
\end{description}

\vspace{-0.1in}
\begin{figure}[ht] 
\centerline{\psfig{file=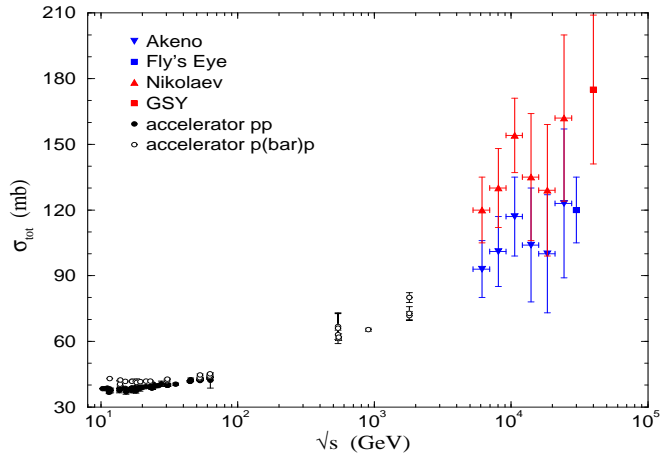, height=7cm,width=10cm}}
\caption{Total cross sections ($pp$ and $\overline{p}p$ above
$\sqrt s = 10$\ GeV): accelerator data and cosmic ray information available.} 
\label{F:lishep1}
\end{figure}

\section{Analytical Approach}	

\subsection{Analyticity relations}

For $pp$ and $\overline{p}p$ scattering, crossing symmetry is expressed by

\begin{eqnarray}  
\sigma_{\pm}(s) = \frac{\sigma_{tot}^{pp} \pm 
\sigma_{tot}^{p\overline{p}}}{2} 
\end{eqnarray} 
and Analyticity allows to connect $\sigma_{tot}(s)$ and $\rho(s)$ through two
compact and symmetric formulas,  

\begin{eqnarray}
\rho ^{pp} \sigma_{tot}^{pp} (s) =
E( \sigma_{+} ) + O( \sigma_{-} ),
\end{eqnarray}

\begin{eqnarray}
\rho ^{p\overline{p}} \sigma_{tot}^{p\overline{p}} (s) =
E( \sigma_{+} ) - O( \sigma_{-} ),
\end{eqnarray}
where, $E( \sigma_{+} )$ and  $O( \sigma_{-} )$ are analytic
transforms relating real and imaginary parts of $E$ven ($+$) and $O$dd
($-$) amplitudes, respectively. They are generally represented by integral
formulas (Integral Dispersion Relations) and in the case of one subtraction
they are expressed by \cite{bc}
 
\begin{eqnarray} 
E_{int}(\sigma_{+}) \equiv
\frac{K}{s} + \frac{2s}{\pi}\int_{s_o}^{\infty} 
\mathrm{d}s' \left[\frac{1}{s'^2-s^2}\right]  
\sigma_{+}(s') = \frac{ReF_{+}(s)}{s}, \nonumber
\end{eqnarray}

\begin{eqnarray} 
O_{int}(\sigma_{-}) \equiv
  \frac{2}{\pi}\int_{s_o}^{\infty} 
\mathrm{d}s' \left[\frac{s'}{s'^2-s^2}\right]  
\sigma_{-}(s') = \frac{ReF_{-}(s)}{s}, \nonumber
\end{eqnarray}
where $K$ is the subtraction constant.

At high energies we can take the limit $s_0 \rightarrow 0$ and in this
case they are equivalent to derivative forms given formally by
\cite{dar,bks,kn,mmp,alm01}

\begin{eqnarray} 
E_{der}(\sigma_{+}) \equiv
\frac{K}{s} + 
\tan\left[\frac{\pi}{2}\frac{d}{d\ln s} \right] 
\sigma_{+}(s) =  \frac{ReF_{+}(s)}{s}, \nonumber
\end{eqnarray}

\begin{eqnarray} 
O_{der}(\sigma_{-}) \equiv
\tan\left[\frac{\pi}{2}( 1 + \frac{d}{d\ln s}) \right] 
\sigma_{-}(s) =  \frac{ReF_{-}(s)}{s}. \nonumber
\end{eqnarray} 
Operationally these expressions may be evaluated through the expansions
\cite{bks,kn,alm01}:

\begin{eqnarray} 
E_{der}(\sigma_{+}) - \frac{K}{s} =
\left[ \frac{\pi}{2} \frac{d}{d\ln s} + 
\frac{1}{3} \left(\frac{\pi}{2}\frac{d}{d \ln s}\right)^3 +
\frac{2}{15} \left(\frac{\pi}{2}\frac{d}{d \ln s}\right)^5 +
...\right] \sigma_{+}(s),
\end{eqnarray}

\begin{eqnarray} 
O_{der}(\sigma_{-})  &=& 
- \frac{2}{\pi}\int \left\{ \frac{d}{d\ln s} \left[\cot \left( \frac{\pi}{2} 
\frac{d}{d\ln s} \right)\right] \sigma_{-}(s) \right\} d\ln s \nonumber \\
&=&
- \frac{2}{\pi}\int \left\{ \left[ 1 - \frac{1}{3}
\left(\frac{\pi}{2}\frac{d}{d \ln s}\right)^2 -  \frac{1}{45}
\left(\frac{\pi}{2}\frac{d}{d \ln s}\right)^4  - ... \right] \sigma_{-}(s)
\right\} d \ln s  \end{eqnarray} 

With this approach, $\rho(s)$ may be determined through suitable input
parametrizations for the total cross sections. In this communication we shall
use two different choices for these inputs:

$\bullet$ Donnachie-Landshoff (DL)

The Donnachie-Landshoff parametrization for the total cross sections
are expressed by \cite{dl}

\begin{eqnarray}
\sigma_{tot}^{pp} (s) = X s^{\epsilon} + Y s^{- \eta}  
\qquad \quad
\sigma_{tot}^{p\overline{p}} (s) = X s^{\epsilon} + Z s^{- \eta} 
\end{eqnarray} 
where, originally, $X = 21.7$ mb, $Y = 56.08$ mb, $Z = 98.39$ mb, 
$\epsilon = 0.0808$, $\eta = 0.4525$.

For this parametrization Eqs. (8) and (9) reads

\begin{eqnarray}
\rho ^{pp} = \frac{1}{\sigma_{tot}^{pp}}
\left\{ \frac{K}{s} +
\left[X\tan(\frac{\pi\epsilon}{2})\right]s^{\epsilon} +
\left[\frac{(Y-Z)}{2}\cot(\frac{\pi\eta}{2}) -
\frac{(Y+Z)}{2}\tan(\frac{\pi\eta}{2})\right]s^{- \eta}\right\} \nonumber
\end{eqnarray}

\begin{eqnarray}
\rho ^{\overline{p}p} = \frac{1}{\sigma_{tot}^{\overline{p}p}}
\left\{ \frac{K}{s} +
\left[X\tan(\frac{\pi\epsilon}{2})\right]s^{\epsilon} +
\left[\frac{(Z-Y)}{2}\cot(\frac{\pi\eta}{2}) -
\frac{(Y+Z)}{2}\tan(\frac{\pi\eta}{2})\right]s^{- \eta}\right\} \nonumber
\end{eqnarray}

We see that this choice implies

\begin{eqnarray}
\Delta \sigma = \sigma^{\overline{p}p}_{tot}(s) - \sigma^{pp}_{tot}(s) =
 (Z - Y) s^{-\eta} \rightarrow 0 \quad {\rm (asymptotically)} \nonumber
\end{eqnarray}

$\bullet$ Kang-Nicolescu (KN)

The parametrizations for the total cross sections under the possibility of a
maximal Odderon contribution are given by \cite{kn}

\begin{eqnarray}
\sigma^{pp}_{tot}(s)= A_{1} + B_{1} \ln s  +
k \ln^2 s, \qquad
\sigma^{\overline{p}p}_{tot}(s)= A_{2} + B_{2} \ln s 
+ k \ln^2 s + 2Rs^{-1/2}
\label{ppbar}
\end{eqnarray}
and Eqs. (8) and (9) give

\begin{eqnarray}
\rho ^{pp} = \frac{1}{\sigma_{tot}^{pp}}
\left\{ \frac{K}{s} + \frac{\pi}{2}\left(\frac{B_1 + B_2}{2}\right) + \left(\pi
k + \frac{A_2 - A_1}{\pi}\right) \ln s + \left(\frac{B_2 - B_1}{2\pi} \right)
\ln ^2 s  - 2Rs^{-1/2} \right\}\nonumber
\end{eqnarray}

\begin{eqnarray}
\rho ^{\overline{p}p} = \frac{1}{\sigma_{tot}^{\overline{p}p}}
\left\{ \frac{K}{s} + \frac{\pi}{2}\left(\frac{B_1 +
B_2}{2}\right) + \left(\pi k - \frac{A_2 - A_1}{\pi}\right)\ln s -
\left(\frac{B_2 - B_1}{2\pi} \right) \ln ^2 s  \right\} \nonumber
\end{eqnarray}

In this case we have

\begin{eqnarray}
\Delta \sigma = \sigma^{\overline{p}p}_{tot}(s) - \sigma^{pp}_{tot}(s) =
(A_2 - A_1) + (B_2 - B_1)\ln s + 2Rs^{-1/2} \rightarrow \Delta A + 
\Delta B \ln s  \nonumber
\end{eqnarray}
so that if $\Delta A
\not= 0$ and/or  $\Delta B \not= 0$ then, asymptotically, $\Delta \sigma \not=
0$.

\section{Fits and Results}

As referred to before, we shall consider several cases: for both
parametrizations (DL and KN) we perform independent and simultaneous fits to
$\sigma_{tot}(s)$ and $\rho(s)$ by considering the two ensembles I
and II, separately.

In the individual procedure we first fit the $\sigma_{tot}$ data through Eqs.
(10) (DL) or (11) (KN) and then fit the $\rho$ data by letting free the
subtraction constant $K$ only; in the simultaneous procedure the subtraction
constant $K$ is a free parameter in the overall fit to both quantities. The
fits were performed through the CERN-minuit routine.

The results with the DL parametrization are shown in Figs. 2 and 3 and the
statistical information are displayed in Table 1. Those with the KN
parametrization in Figs. 4 and 5 and Table 2.

\section{Conclusions and final Remarks}

From  Tables 1 and 2 and Figures 2 to 5 we are lead to the following main
conclusions:

$\bullet$ Ensembles I and II

In general, the results with ensembles I and II do not differ
substantially and the cosmic ray information in ensemble II (Nikolaev and GSY)
are not described in all the cases. Certainly this is due to the small number
of points and the large error bars as compared with the accelerator data and
also to the choices for the parametrizations. However noticeable differences
occur with the Kang Nicolescu parametrization (Figures 4 and 5).

\vspace{0.3cm}

$\bullet$ Simultaneous and individual fits

The best statistical results are obtained through the simultaneous fits.
For example, in the case of  DL parametrization, $\chi^2 / d.o.f.
\sim 1.0$ (simultaneous) and $\sim 1.7$ (individual for $\rho$).
This is expected since in the simultaneous case we have more statistical
information (degree of freedom).

On the other hand simultaneous fits constraint the rise of the total cross
section. For example, in the case of the DL parametrization with ensembles I
and II we obtained:

\vspace{0.2cm}

\centerline{ $\epsilon_{indi}^{I} \sim 0.088 \rightarrow 
\epsilon_{simul}^{I} \sim 0.083$ (reduction $\sim 6 \% $)}

\vspace{0.2cm}

\centerline{$\epsilon_{indi}^{II} \sim 0.091 \rightarrow
\epsilon_{simul}^{II} \sim 0.084$ (reduction $\sim 8 \% $)}

\vspace{0.2cm}

As commented before, $\sigma_{tot}$ and $\rho$ do not have the same status as
physical quantities since the $\rho$ value is model dependent. Therefore
we understand that the simultaneous fit may led to erroneous predictions 
for $\sigma_{tot}$ in
the asymptotic region (underestimation).

\vspace{0.3cm}

$\bullet$ Upper bounds for the soft Pomeron intercept

The upper bound was obtained with ensemble II in the case of
individual fits

\centerline{$1 + \epsilon = 1.091 \pm 0.003 \qquad 1 - \eta = 0.65 \pm 0.03$}

\vspace{0.3cm}

$\bullet$ Odderon

From Tables 3 and 4, $\Delta A$ and $\Delta B$ $\sim$ $0$ in 3 cases
and the only exception is for individual fits with ensemble II,
for which $\Delta A = 5 \pm 3$ and $\Delta B \sim 0$. This predicts
a crossing with $\sigma_{tot}^{pp}$ becoming higher than
$\sigma_{tot}^{\overline{p}p}$. However, in this case
$\rho$ is not described as shown in Fig. 5. We conclude that our
results do not favor the Odderon possibility.

We observe that a similar and interesting analysis, taking quantitative account
of the cosmic ray information, has been presented in this Workshop by Carlos
Avila \cite{avila}. Although the technical approach is different the global
results are in agreement with those presented here and in Ref. \cite{alm01}.

\vspace{0.5cm}
\leftline{\bf Acknowledgments}
Work supported by FAPESP and CNPq, Brazil. M.J.M.
is thankful to Prof. Alberto Santoro and the organizers of this
excellent Workshop and to U. Maor and C. Avila for useful discussions
during the meeting.

\newpage
\begin{minipage}{11.0cm}
\begin{table}[ht!]
\caption{Simultaneous and individual fits to $\sigma_{tot}$ and
$\rho$ for Ensembles I and II with the DL
parametrization.} 
\label{F:dl}
\begin{tabular}{||c||c|c||c|c|| } \tableline \tableline
Fit: & \multicolumn{2}{c||}{Simultaneous} & \multicolumn{2}{c||}{Individual
} \\ \tableline 
Ensemble:  &  I &  II &  I &  II  \\ \tableline \tableline
N. points  &   $165$    &   $165$ & $102$ & $102$ \\ \tableline  
$\chi^2/d.o.f.$   &  $0.84$ &  $0.98$ & $0.76$ & $0.96$  \\ \tableline     
$X$ (mb) & $ 21.4 \pm 0.4$  & $21.1 \pm 0.4$ & $20.0 \pm 0.7$ & $ 19.3 \pm
0.7$   \\ \tableline   
$Y$ (mb) & $ 67 \pm 6$  & $67 \pm 5$ & $48 \pm 5 $ & $46 \pm 5$  \\
\tableline   
$Z$ (mb) & $114 \pm 11$ & $112 \pm 11$ & $75 \pm 10$ & $70 \pm 9$  \\
\tableline   
$\eta$  & $0.48 \pm 0.02$ &  $0.47 \pm 0.02$ & $0.37 \pm 0.04$ & $0.35
\pm 0.03$ \\ \tableline 
$\epsilon$ & $0.083 \pm 0.002$  &  $0.084 \pm 0.002$ & $0.088 \pm 0.003$ &
$0.091 \pm 0.003$ \\ \tableline   
$K$ & $306 \pm 54$  & $307 \pm 54$ & - & -   \\  \tableline \tableline 
N. points ($\rho$) & - & - &   $63$     &    $63$  \\ \tableline  
 $\chi^2/d.o.f.$ & - & - & $1.55 $ &  $ 1.84 $  \\ \tableline        
$K$ & - & - & $235 \pm 32$  &   $245 \pm 32$    \\ \tableline 
\end{tabular}
\end{table}
\end{minipage}


\begin{minipage}{11.0cm}
\begin{table}[ht!]
\caption{Simultaneous and individual fits to $\sigma_{tot}$ and
$\rho$ for Ensembles I and II with the KN parametrization.} 
\label{F:kn}
\begin{tabular}{||c||c|c||c|c|| } \tableline \tableline
Fit: & \multicolumn{2}{c||}{Simultaneous} & \multicolumn{2}{c||}{Individual}
\\ \tableline  Ensemble:  &  I &  II &  I &  II  \\ \tableline \tableline
N. points  &    165   &   165  & 102 & 102\\ \tableline  
$\chi^2/d.o.f.$& $0.77 $ &  $0.86$ & $0.78$ & $0.91$   \\ \tableline        
$A_1$ (mb) & $ 44.4 \pm 0.7$  & $45.0 \pm 0.7$ & $ 45 \pm 1$ & $ 47 \pm 1$   \\
\tableline   
$A_2$ (mb) & $ 44.5 \pm 0.7$  & $45.1 \pm 0.7$ & $45 \pm 3$ & $52 \pm 3$ \\
\tableline   
$B_1$ (mb)  & $ -2.9 \pm 0.2   $  &  $ -3.1 \pm 0.2$ & $-2.9 \pm 0.3$ & $-3.6
\pm 0.4$  \\ \tableline   
$B_2$ (mb)  & $ -2.9 \pm 0.2 $  &  $ -3.1  \pm 0.2$ & $-2.9 \pm 0.6$ & $-4.2
\pm 0.6$ \\ \tableline 
$k$ (mb) & $  0.33 \pm 0.01$  &  $ 0.35 \pm 0.01$ & $0.33 \pm 0.06$ & $0.39
\pm 0.03$  \\ \tableline  
$R$ (mb\ GeV) & $ 25.3 \pm 0.9$  & $25.4 \pm 0.9$ & $24 \pm 7$ & $12 \pm 7$  
\\ \tableline        
$K$ & $ -72 \pm 46$  & $-63 \pm 46$ & - & -  \\ \tableline \tableline  
N. points ($\rho$) & - & -   &   63     &    63  \\ \tableline  
$\chi^2/d.o.f.$& - & - &  $1.05 $ &  $ 35.2 $  \\ \tableline        
$K$ & - & - & $ -198 \pm 32$  & $-1369 \pm 32  $    \\ \tableline 
\end{tabular}
\end{table}
\end{minipage}

\newpage

\clearpage

\vspace{-0.1in}
\begin{figure}[ht] 
\centerline{\psfig{file=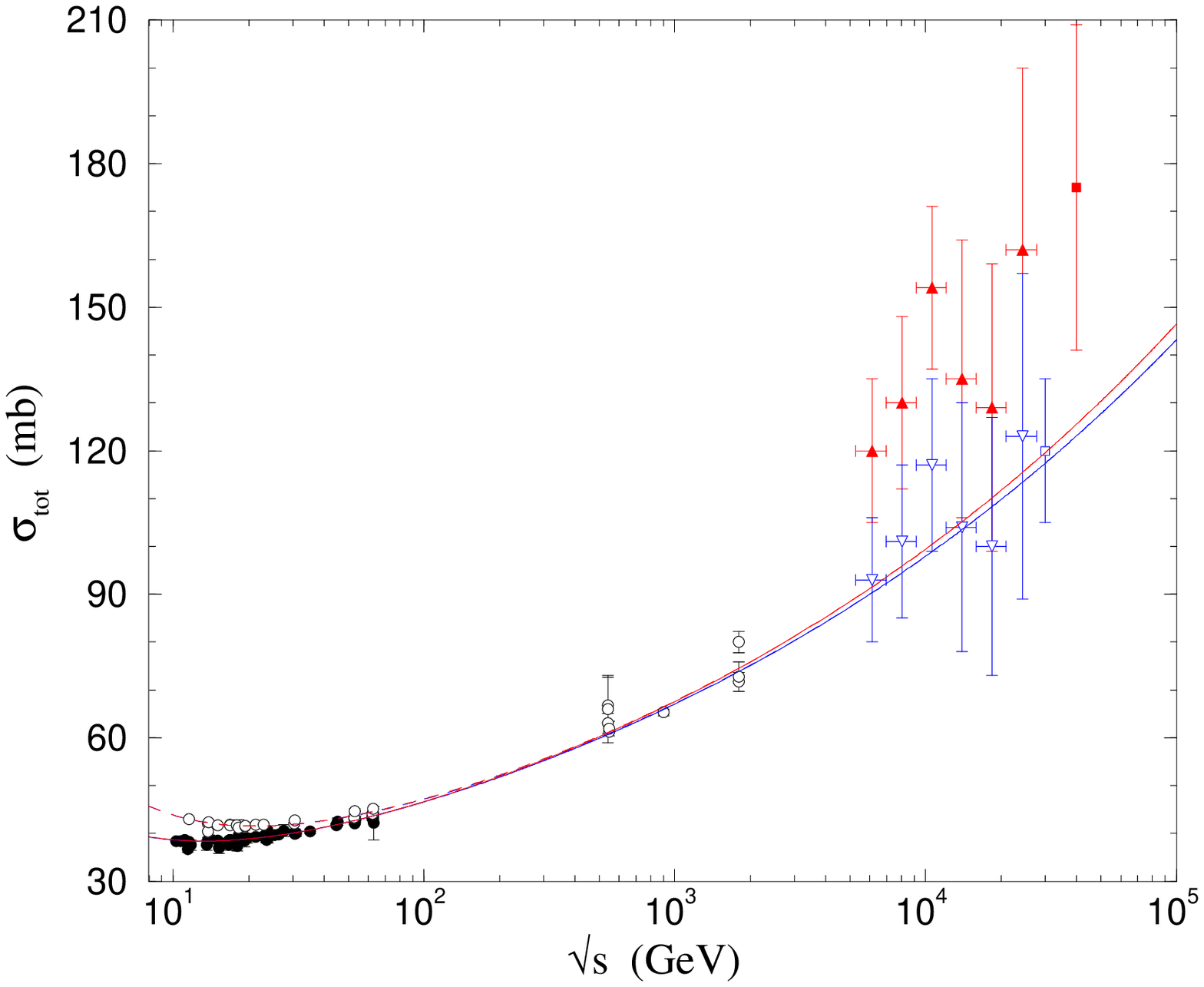, height=10cm,width=15cm}}
\centerline{\psfig{file=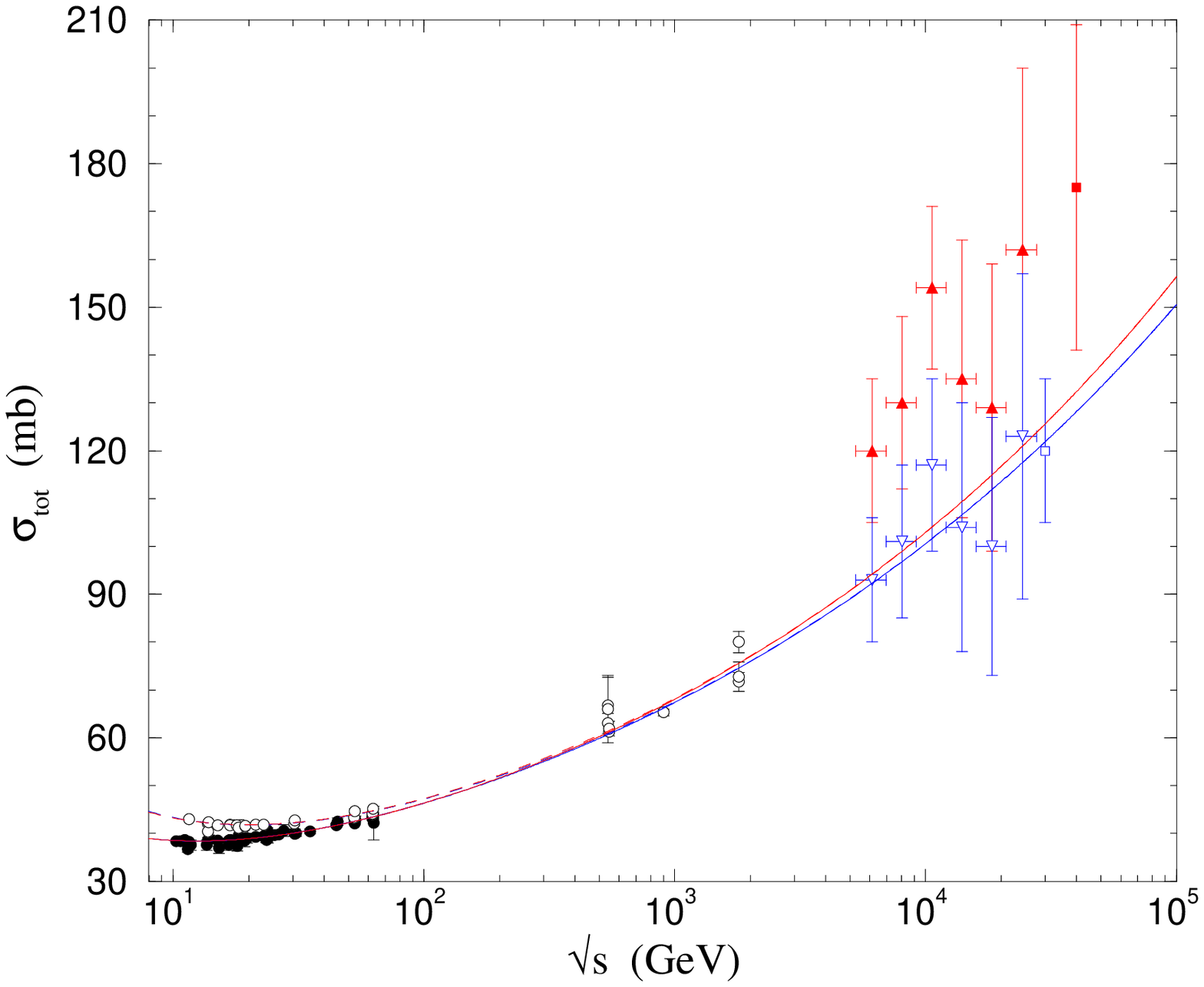, height=10cm,width=15cm}}
\caption{Fits to $\sigma_{tot}$  with the
DL parametrization: 
 $pp$ (solid) and $\overline{p}p$ (dashed) and ensembles I (blue) and II
(red). Simultaneous (up) and individual (down).} 
\label{F:2lishep}  
\end{figure}

\vspace{-0.1in}
\begin{figure}[ht] 
\centerline{\psfig{file=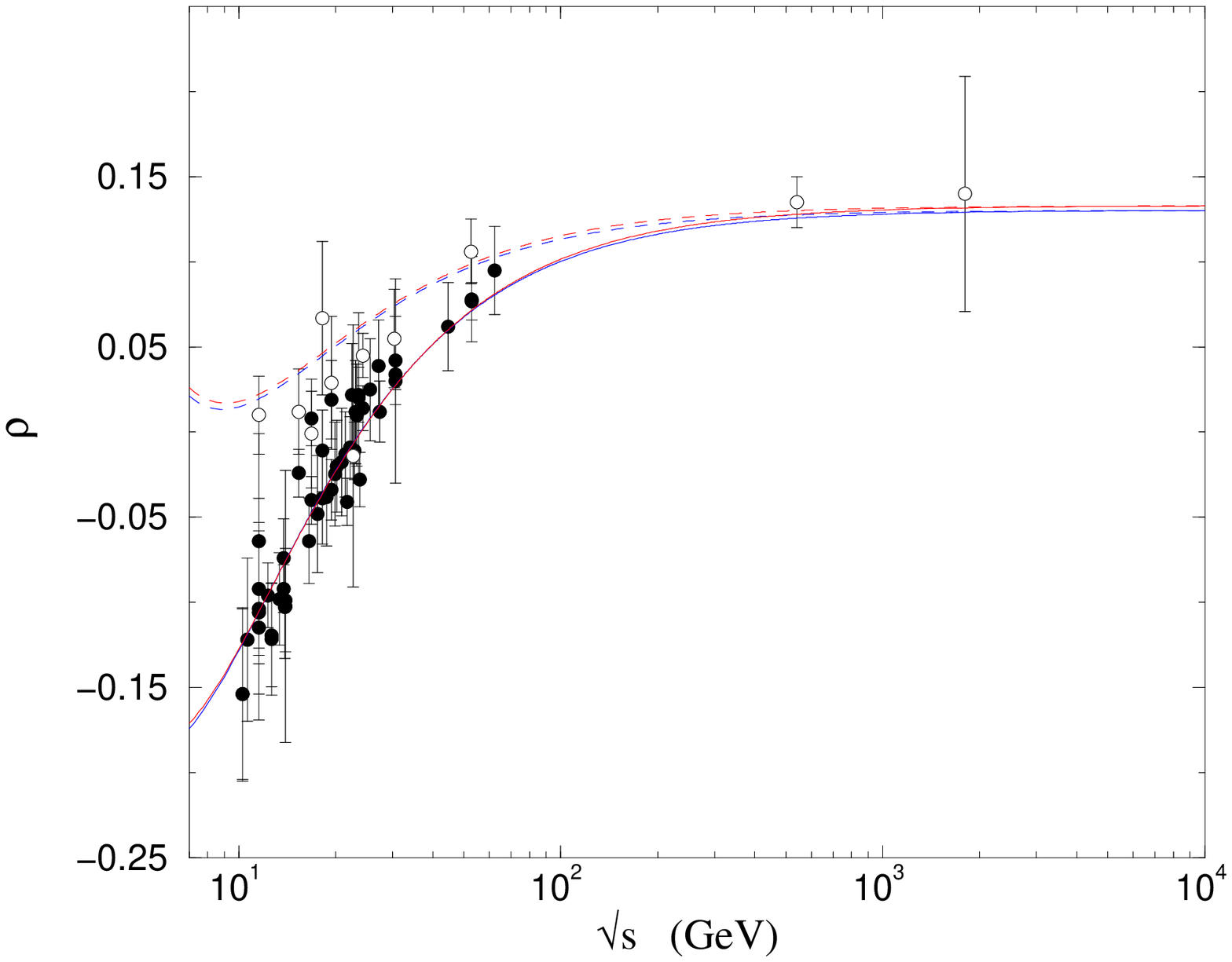, height=10cm,width=15cm}}
\centerline{\psfig{file=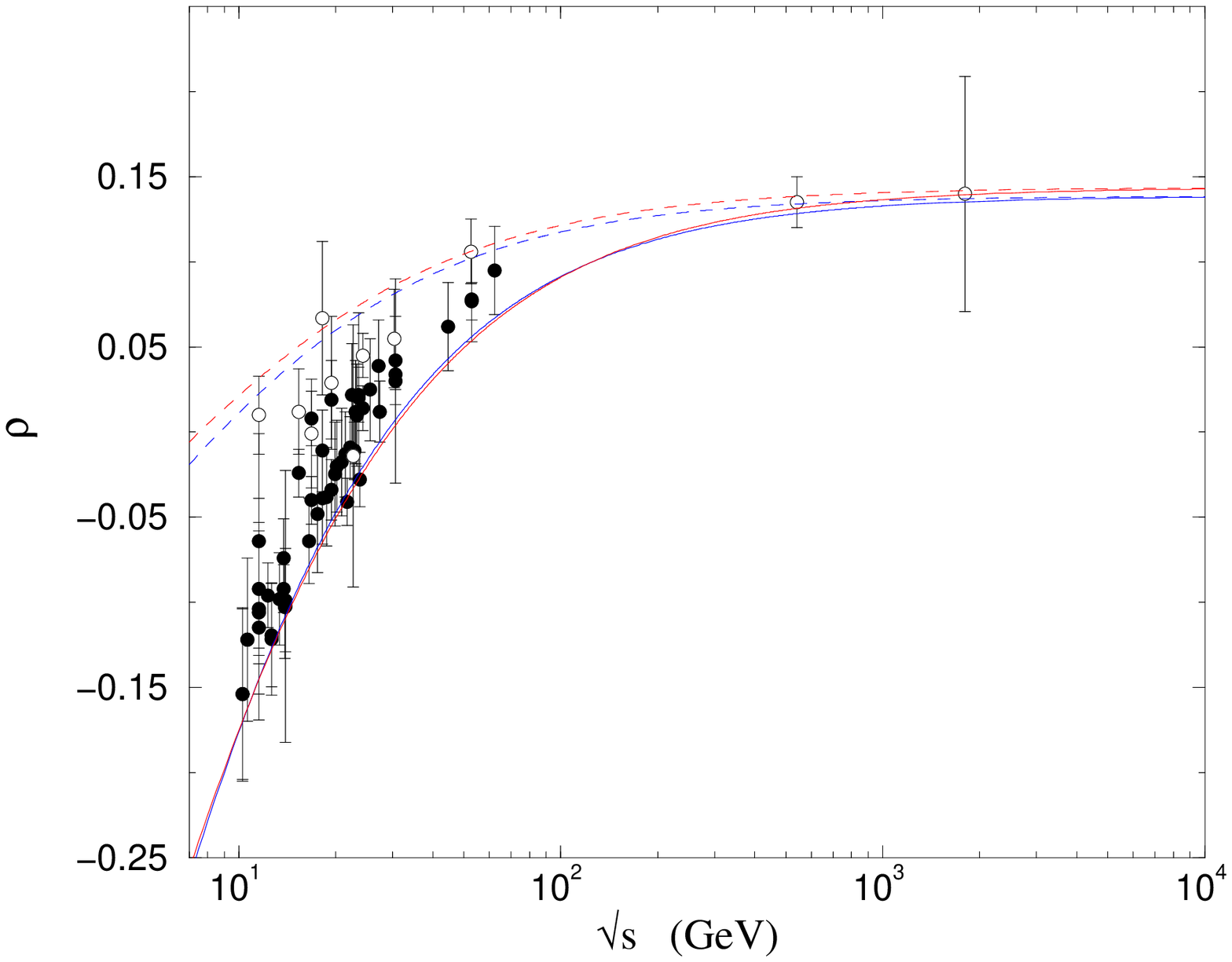, height=10cm,width=15cm}}
\caption{Fit to  $\rho$ with the
DL parametrization: 
 $pp$ (solid) and $\overline{p}p$ (dashed) and ensembles I (blue) and II
(red). Simultaneous (up) and individual (down).} 
\label{F:3lishep}  
\end{figure}

\vspace{-0.1in}
\begin{figure}[ht] 
\centerline{\psfig{file=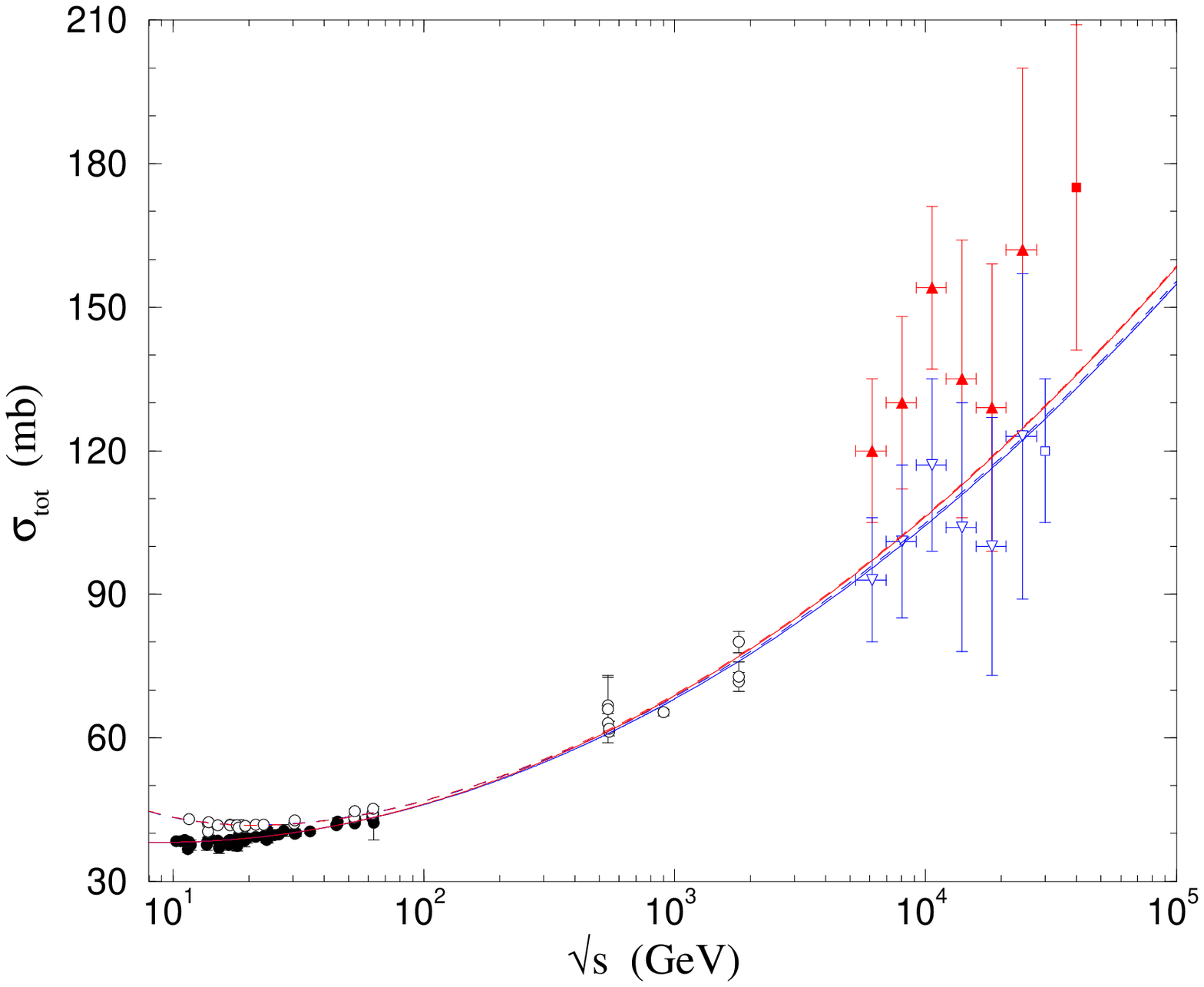, height=10cm,width=15cm}}
\centerline{\psfig{file=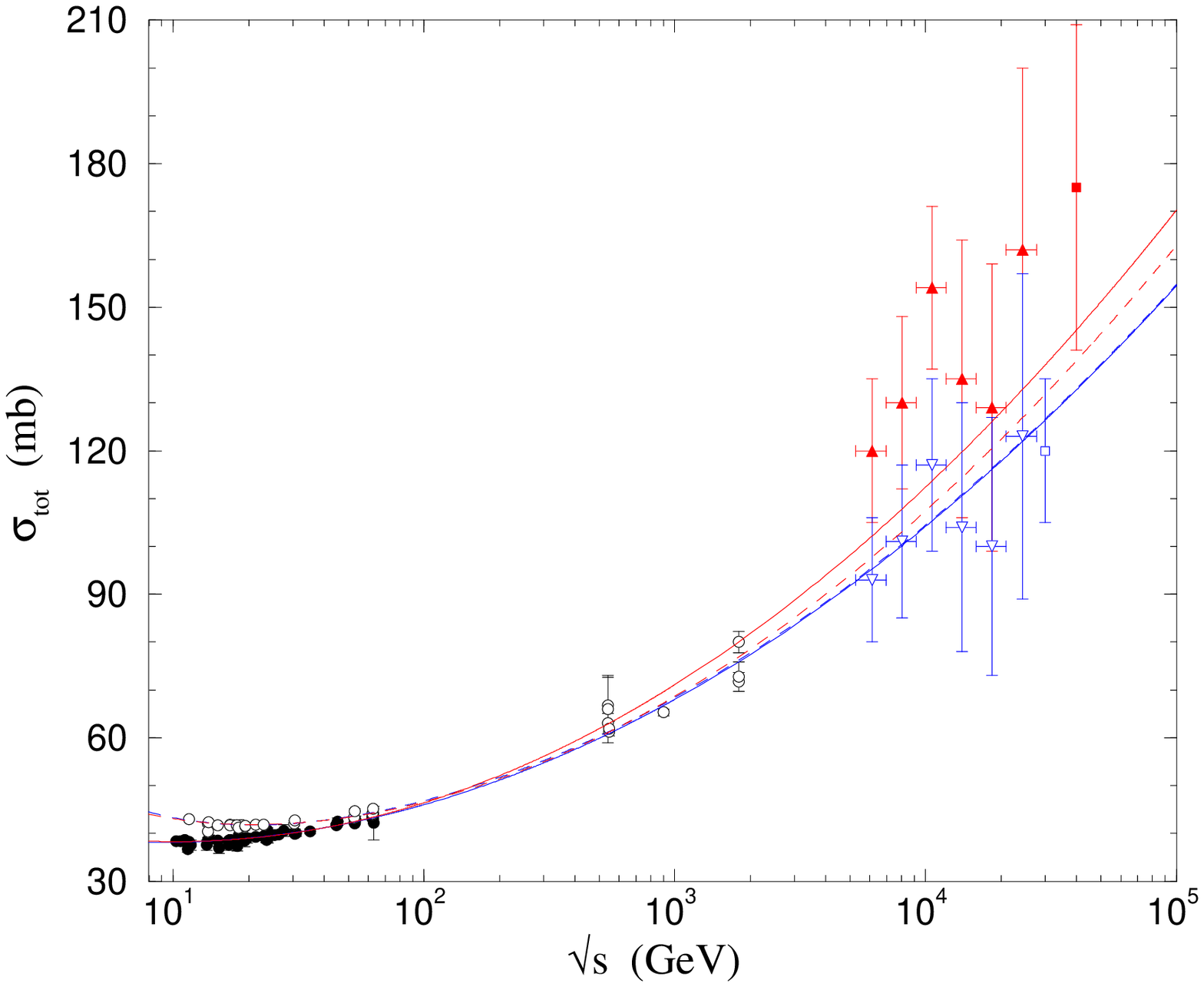, height=10cm,width=15cm}}
\caption{Fits to $\sigma_{tot}$  with the
KN parametrization: 
 $pp$ (solid) and $\overline{p}p$ (dashed) and ensembles I (blue) and II
(red). Simultaneous (up) and individual (down).} 
\label{F:4lishep}  
\end{figure}

\vspace{-0.1in}
\begin{figure}[ht] 
\centerline{\psfig{file=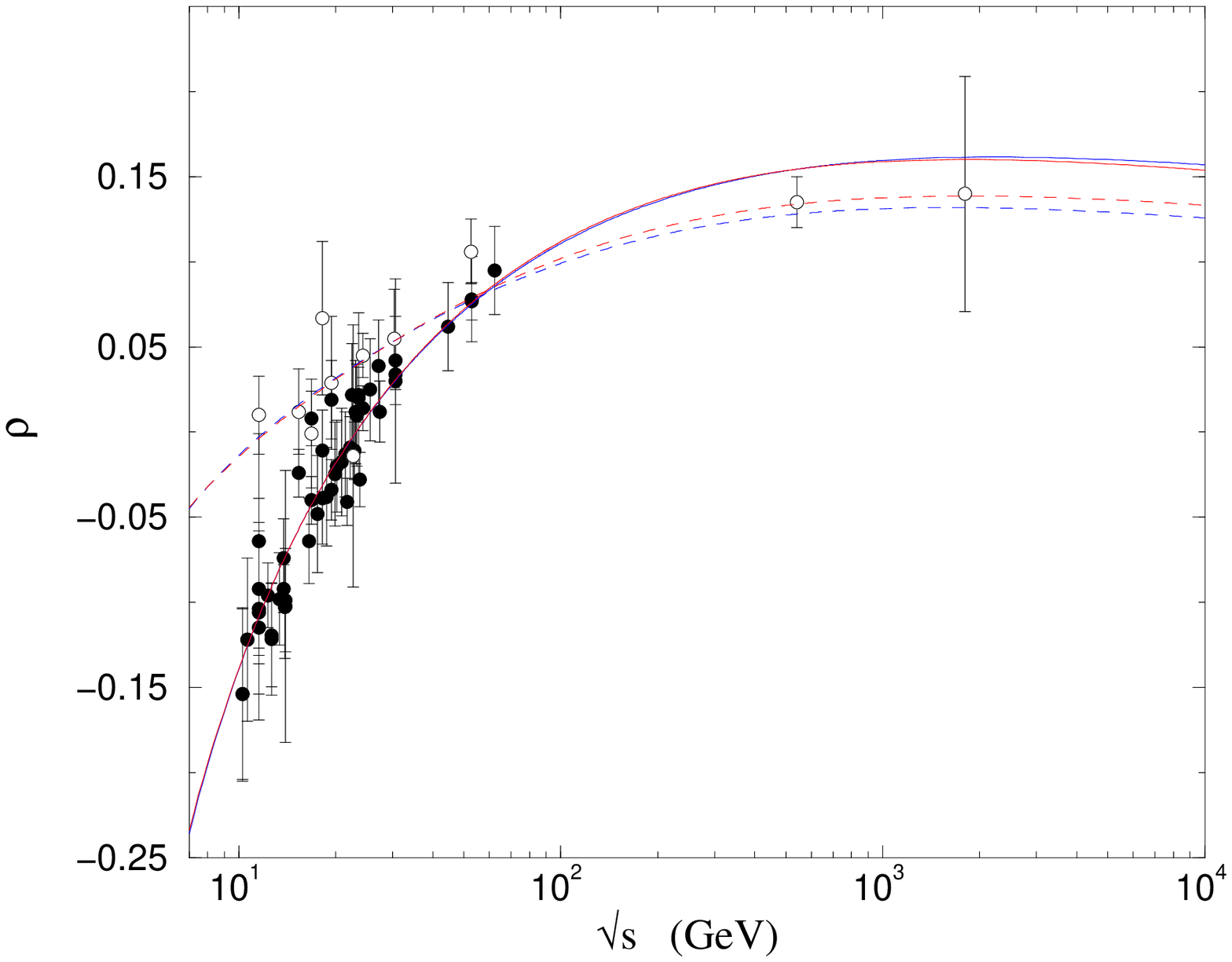, height=10cm,width=15cm}}
\centerline{\psfig{file=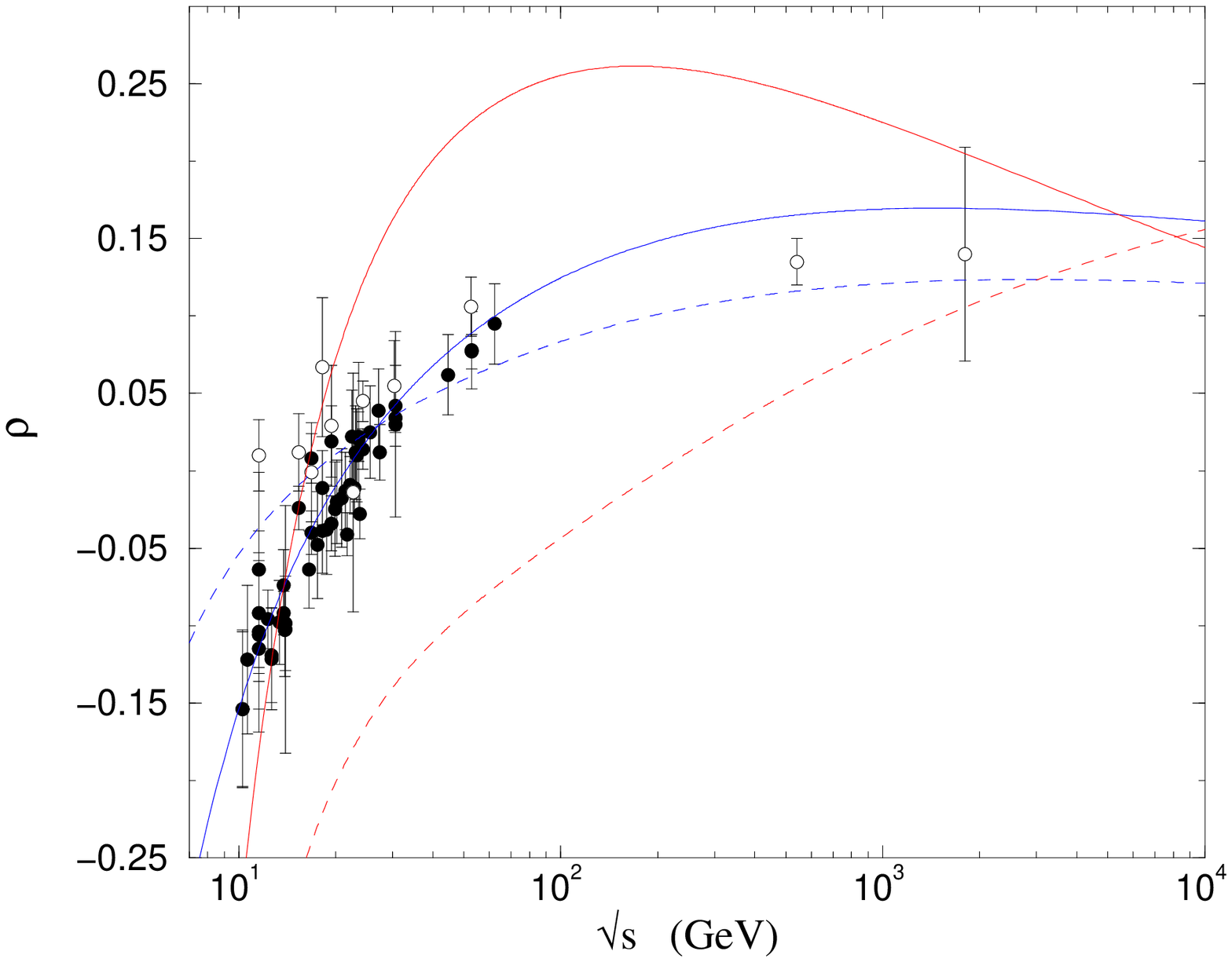, height=10cm,width=15cm}}
\caption{Fit to $\rho$ with the
KN parametrization: 
 $pp$ (solid) and $\overline{p}p$ (dashed) and ensembles I (blue) and II
(red). Simultaneous (up) and individual (down).} 
\label{F:5lishep}  
\end{figure}

\end{document}